# An indirect measurement protocol of intracavity mode quadratures dispersion in Dynamical Casimir Effect

[1]Miroshnichenko G.P., [2]Trifanova E.S., [3]Trifanov A.I.
gpmirosh@gmail.com, etrifanova@gmail.com, alextrifanov@gmail.com

ITMO University, 197101 Kronverkskyavenue 49, Saint Petersburg, Russia

**Abstract:** This work deals with the problem of photon detection generated from the mirror-induced Dynamical Casimir Effect. Particularly we are interested in measurement of those statistical characteristics of a generated intracavity field which may confirm its nonthermal properties. Here an indirect protocol for quadrature dispersion measurement is presented.

**1. Introduction**

Vacuum fluctuations being predicted by Quantum Electrodynamics is one of the fundamental effects which lie in the background of our understanding of Quantum Physics. At the beginning of 20-th century the problem of vacuum renormalization gave the prediction of the Casimir Effect which was successfully observed in laboratory not long ago [1]. In 1970s several effects connected to vacuum fluctuations were predicted [2, 3, 4]. One of these phenomena was the Dynamical Casimir Effect (DCE) [4, 5] – the process of real photon creation out of vacuum in a cavity due to the accelerated motion of its boundary [6]. In the past four decades numerous works devoted to investigations of spectral and statistical properties of the Casimir field [7, 8, 9, 10, 11] and considerations of experimental verification possibility [12, 13, 14, 15] were published.

Particularly, it was shown that the motion of cavity walls is accompanied by squeezing of the initially vacuum state of the field inside and an intermode interaction takes place [7, 8]. Besides a pure fundamental interest, this effect could have useful applications for generation of the parametrical radiation. Due to the absence of possibility to organize mechanical motion of cavity boundaries with a very high frequency different ideas of using superconducting [15, 16] and semiconducting [17, 18] systems were considered. Also, changing in material properties of the intracavity medium was suggested [19, 20].

In 2011 it was reported about a successful experiment on observation of DCE in a superconducting circuit [21, 22]. Using superconductive device (SQUID) the authors realized the imitation of the boundary motion and DCE field generation in GHz frequency range. These works initiated a discussion about the most effective measurement protocol of created quanta detection and modification of the intracavity field statistics due to the interaction with the measurement apparatus [23]. Particularly, different models for the measurement apparatus from quantum optics were suggested and thoroughly investigated [24, 25, 26]. It should be noted that DCE field state detection assumes measurement of those statistical characteristics which confirm its nonthermal character, namely:quadrature squeezing, correlation functions, etc.

It is worthwhile to recall that in GHz frequency range application of a standard cavity QED measurement technique [27] with microscopic atom-pointer detectors in use[23, 28] is available. And furthermore, superconducting circuits allow realization systems with an ultrastrong atom-cavity coupling [29]. Using these facts we awarded to suggest the indirect measurement protocol (IMP) for detection of the Casimir field statistical properties.

A standard indirect photodetection protocol consists of the following steps. An atom, prepared in its ground state, passes through the cavity and interacts with the excited mode of the intracavity field. Just after this interaction the atom interacts with the additional mode of the electromagnetic field and then goes to the atomic state detector. Using statistics of the detector results one can obtain statistical properties of the intracavity field. This detection technique gives rich facilities in realizing weak and unsharp measurements [30, 31, 32], accounting imperfectness of measurement devices [33, 34, 35].

Here we propose the general indirect measurement protocol for investigation of statistical properties of the Casimir field. Particularly, we are interested in quadrature dispersion of the intracavity mode to check nonclassical properties of the field and prove its nonthermal character. The structure of our work is as follows: first of all we recall the dynamical properties of the system consisting of the cavity with moving wall(s) and write Hamiltonian which governs the evolution of the quantum state of the intracavity field. In section 3 a general protocol of the indirect quantum measurement will be described and its application to the problem of the field state detection in DCE will be suggested. In section 4 basic ideas of quadrature dispersion measurement will be introduced and applied to state the squeezing detection. Section 5 contains numerical modeling and discussion. Section 6 concludes the paper.

## 2. Effective Hamiltonian for DCE

Dynamical Casimir Effect is a relativistic process of photon creation out of vacuum due to the accelerated motion of the cavity mirrors or changing of the intracavity media electromagnetic properties. Its dynamics may be described by effective Hamiltonian, which consists of three terms describing the effects of free evolution, squeezing and intermode interaction [7, 9] (here and further $\hbar = 1$):

$$H_{eff} = \sum_k \omega_k(t) a_k^\dagger a_k + i \sum_k \xi_k(t)\left(a_k^{\dagger 2} - a_k^2\right) + \frac{i}{2} \sum_{j \neq k} \mu_{kj}(t)\left(a_k^\dagger a_j^\dagger + a_k^\dagger a_j - a_j a_k - a_j^\dagger a_k\right). \quad (1)$$

Here $a_k$, $a_k^\dagger$ are annihilation and creation operators for $k$-th mode, $\omega_k(t)$ is instantaneous frequency, $\xi_k(t)$ and $\mu_{kj}(t)$ are coefficients which depend on geometrical properties of the cavity and on the law of boundary motion.

As it was mentioned above we will concentrate on the parameters which characterize squeezing properties of the field state. So for the following simplification we assume conditions which allow one mode approximation with squeezing as a main source of Casimir light in particular mode. For the case of spatially homogeneous dielectric permittivity $\varepsilon(x,t) = \varepsilon(t)$ without boundary motion, one can obtain $\mu_{kj}(t) = 0$ and one-mode approximation is applicable [7]:

$$H_C = \omega_F(t) a^\dagger a + i\xi(t)\left(a^{\dagger 2} - a^2\right) = H_0 + H_{sq}, \quad (2)$$

$$\xi(t) = \frac{1}{4\omega_F(t)} \frac{d\omega_F(t)}{dt}. \quad (3)$$

For the following, our goal is to present the quadrature squeezing measurement protocol of DCE for the degenerate regime of photon creation. It should be noted that another way to measure these characteristics is to organize the protocol of the homodyne detection, which requires the amplification of the detected Casimir field. We suggest the effective indirect measurement protocol without amplification.

## 3. General indirect photodetection protocol for DCE

Here we introduce the algorithm for measurement of the Casimir field state statistical properties. We start from the Casimir field generation from vacuum state $\rho_{vac}^F$ in QED cavity due to DCE(2). Then a two-level atom initially prepared in its ground state $\rho_0^A = |g\rangle\langle g|$ goes through this cavity and interacts with the mode monitoring statistical properties of its resulting state $\rho_{in}^F$. Common evolution of the atom-mode system is described by the Hamiltonian:

$$H_{a-f} = H_C + \frac{\omega_a}{2}\sigma_z + g\left(a^\dagger \sigma_- + a\sigma_+\right) = H_C + H_A + H_{int}, \qquad (4)$$

where $H_C$ is taken from (2), $\omega_a$ is atomic frequency, $\sigma_z = |e\rangle\langle e| - |g\rangle\langle g|$, $\sigma_- = |g\rangle\langle e|$, $\sigma_+ = |e\rangle\langle g|$ are atomic operators and $g$ is a coupling parameter.

Stopping moving the wall just before the beginning of the measurement process is undesirable since fast oscillation frequency of the free field leads to the meantime values of quadratures which in fact do not correspond to squeezing [19]. Actually, one can easily check that the following quantity

$$\left\langle(\Delta q)^2\right\rangle = \left\langle q^2\right\rangle - \left\langle q\right\rangle^2 = \mathrm{tr}\left[U_0^\dagger \cdot q^2 \cdot U_0 \rho_{sq}\right], \qquad (5)$$

where $U_0 = \exp(-i\omega_f a^\dagger a t)$ is a free evolution operator of the intracavity mode, $\rho_{sq} = S(\xi)\rho_{vac}^F S^\dagger(\xi)$ is the squeezed vacuum state, $\xi = re^{i\varphi}$ is a squeezing parameter and $q$ is either $x = (a + a^\dagger)/\sqrt{2}$ or $p = (a - a^\dagger)/(\sqrt{2}i)$, leads to (upper bar denotes the meantime value of quadrature)

$$\overline{\left\langle(\Delta x)^2\right\rangle} = \overline{\left\langle(\Delta p)^2\right\rangle} = \mathrm{sh}^2(r) + 1/2. \qquad (6)$$

At the next step the atom leaves the cavity and interacts with the external mode of the classical electromagnetic field. Corresponding Hamiltonian then may be written as follows:

$$H_{ext} = H_A + \left(\Omega e^{-i\omega_e t}\sigma_- + \Omega^* e^{i\omega_e t}\sigma_+\right) = H_A + V_{ext}. \qquad (7)$$

Here $\Omega$ is Rabi frequency of the mode, $\omega_e$ is the mode frequency and $\sigma_\pm$ are atomic operators described above. After interaction the atom goes to the detector where its state is determined. We suppose that the atomic state detection procedure is ideal and projective and it is described by the following set of projector-valued measures:

$$\Pi_r^A = |r\rangle\langle r|, \; r \in \{g, e\}. \qquad (8)$$

The probability of correspondent detection result $r$ may be calculated as

$$P_r = \mathrm{Tr}\left(\rho_{AF}\Pi_r^A\right), \qquad (9)$$

where $\text{Tr}(...) = \text{tr}_A \text{tr}_F(...)$ is a full trace over field and atomic state spaces and $\rho_{AF}$ is a common atom-field state just before the atomic state detection. Using this expression we can introduce the nonorthogonal set of the field state detector measures:

$$P_r = \text{tr}_F\left(\text{tr}_A\left(\rho_{AF}\Pi_r^A\right)\right) = \text{tr}_F\left(\rho_{in}^F\Pi_r^F\right) = \text{tr}_F\left(\rho_r^F\right), \qquad (10)$$

where $\rho_r^F$ is the intracavity field state conditioned by the atomic state detection result.

**4. Conditional evolution for Casimir field**

To describe the idea of quadrature squeezing measurement we recall that the final state of the field is conditional since it depends on the detection result. It may be evaluated using the set of Kraus transformers $\{K_r\}$, which may be introduced in the following way[36]:

$$\text{tr}_F\left(\rho_{in}^F \cdot \Pi_r^F\right) = \text{tr}_F\left(\rho_r^F\right) = \text{tr}_F\left(K_r \rho_{in}^F K_r^\dagger\right) \qquad (11)$$

They act on the initial state of the field and map it onto the conditional final state (first Kraus representation theorem):

$$\rho_{in}^F \to \rho_r^F = \frac{K_r \rho_{in}^F K_r^\dagger}{\text{tr}_F\left(K_r \rho_{in}^F K_r^\dagger\right)}, \quad r \in \{g,e\} \qquad (12)$$

Here $K_g$ corresponds to the state transformation for the case when the atom was detected in its ground state $|g\rangle$ and $K_e$ corresponds for the case of the excited state detection. Now we can calculate corresponding probabilities for these events as

$$P_r^{(1)} = \left\langle K_r^\dagger K_r \right\rangle_{in}, \quad r \in \{g,e\}. \qquad (13)$$

For the following we will also be interested in probabilities to detect two subsequent atoms in the same state after interaction lasting time $\tau$ in each detection cycle:

$$P_{rr}^{(2)} = \left\langle K_r^\dagger K_r^\dagger K_r K_r \right\rangle_{in}, \quad r \in \{g,e\}. \qquad (14)$$

To get some analytical formulas we write common evolution of the atom and intracavity field in weak interaction approximation,

$$\tau\left\langle a^\dagger a \right\rangle = g t_{int} \left\langle a^\dagger a \right\rangle \ll 1. \qquad (15)$$

This may be realized either through the small interaction time $t_{int}$ or small coupling parameter $g$. Using the first condition we can neglect the field state change due to the DCE during the atom-field interaction. Then Kraus transformers may be written as those in Jaynes-Cummings model,

$$\begin{aligned} K_g &= \cos\left(\tau\sqrt{a^\dagger a}\right), \\ K_e &= -i\, a \sin\left(\tau\sqrt{a^\dagger a}\right)\Big/\sqrt{a^\dagger a}. \end{aligned} \qquad (16)$$

Using them for calculating the probability to detect the atom in its ground state one can obtain

$$P_g^{(1)} = \left\langle K_g^\dagger K_g \right\rangle = \left\langle 1 - \tau^2 a^\dagger a + ... \right\rangle, \qquad (17)$$

where we used expansion $P_g^{(1)}(\tau)$ up to the second order in $\tau$. This expression, for example, may be utilized to find the mean photon number in the cavity. Namely, using statistics of the ground state detector clicks the following expression takes place:

$$\langle n \rangle = \langle a^\dagger a \rangle = \frac{d^2}{d\tau^2} P_g^{(1)}(0). \tag{18}$$

After that we organizethe interaction between the atom and the mode of the classical electromagnetic field. Using $\Omega = \pi/2$ in the interaction picture the following transformation of the atomic state is realized

$$U_{ext} = \exp\left[-i\frac{\pi}{4}\left(\sigma_- e^{i\theta} + \sigma_+ e^{-i\theta}\right)\right], \tag{19}$$

where $\theta = (\omega_e - \omega_a)T$ and $T$ is the atom-mode interaction time. This atomic state transformation corresponds to the following transformation for Kraus operators:

$$K_g \to \tilde{K}_g = \frac{1}{\sqrt{2}}\left[K_g + e^{-i\theta}K_e\right], \quad K_e \to \tilde{K}_e = \frac{1}{\sqrt{2}}\left[e^{i\theta}K_g + K_e\right]. \tag{20}$$

Then the probabilities $\tilde{P}_g^{(1)}(\tau)$ and $\tilde{P}_g^{(2)}(\tau)$ to detect one atom and two subsequent atoms in the ground state may be expanded in series up to the second order of $\tau$

$$\tilde{P}_g^{(1)} = \langle \tilde{K}_g^\dagger \tilde{K}_g \rangle = \frac{1}{2}\langle 1 + \sqrt{2}\tau Q + ...\rangle, \tag{21}$$

$$\tilde{P}_g^{(2)} = \langle \tilde{K}_g^\dagger \tilde{K}_g^\dagger \tilde{K}_g \tilde{K}_g \rangle = \frac{1}{4}\langle 1 + 2\sqrt{2}\tau Q + \tau^2(2Q^2 - 1) + ...\rangle. \tag{22}$$

Under the known atomic state detection statistics this gives mean values for operator $Q = \dfrac{e^{-i\theta}a - e^{i\theta}a^\dagger}{\sqrt{2}i}$ and $Q^2$, which in fact are proportional to the desired quadratures:

$$\langle Q \rangle = \sqrt{2}\frac{d}{d\tau}\tilde{P}_g^{(1)}(0), \tag{23}$$

$$\langle Q^2 \rangle = \frac{d^2}{d\tau^2}\tilde{P}_g^{(2)}(0) + \frac{1}{2}. \tag{24}$$

## 5. Numerical simulation

In our numerical scheme we assumed a periodical regime of frequency modulation, which corresponds to the exponential growth of photon number in the cavity [28]:

$$\omega_F(t) = \omega_0(1 + \varepsilon \sin \eta t). \tag{25}$$

Here $\eta = 2\omega_0$ (condition of exponentially growth of photon number) and $\varepsilon$ is the modulation depth.Also we chose the following parameters:

$$\Delta = 0, \quad g = 0.5 \times 10^{-3}, \quad \varepsilon\eta = 2 \times 10^{-3}, \tag{26}$$

where $\Delta$ is the atom-cavity mode frequency detuning.

In Fig. 1 basic statistical characteristics as functions of the interaction time are presented. Two cases are under consideration: an exact solution for the atom-mode interaction followed by DCE (solid line) and an approximate solution when DCE process is dropped out during the intracavity interaction (dashed line).

The following procedure is simulated. In the time interval from 0 to 1 mcs field generation due to the DCE takes place. Here there is no interaction between the atom and the mode (Fig.1a), the mean intracavity photon number increases exponentially (Fig.1b) and one of the quadratures is being squeezed (Fig.1c).

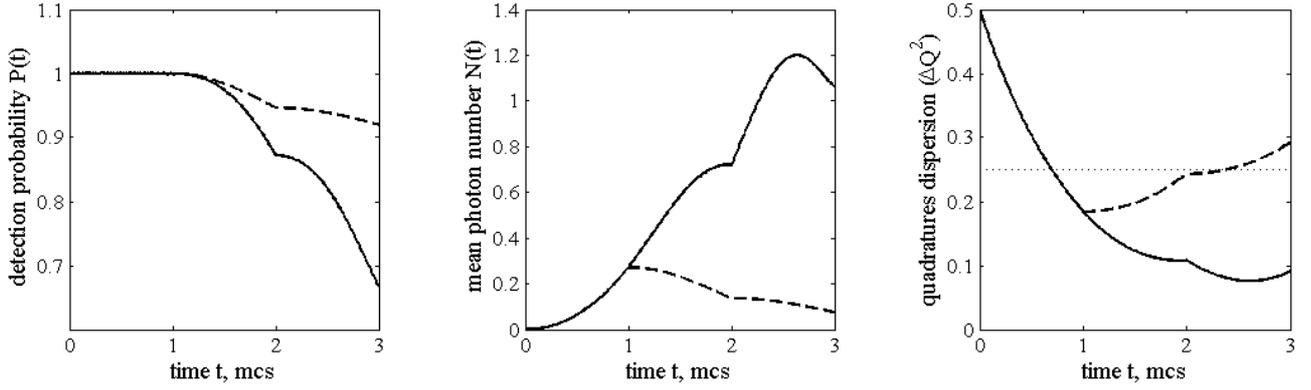

**Fig. 1** Basic statistical characteristics of DCE field conditional state: probability of the atomic ground state detection (a), mean photon number (b) and quadrature X dispersion (c) as functions of interaction time. Two cases are under consideration: with (solid line) and without (dashed line) trembling the boundary during the atom-field interaction

In the next time interval from 1 to 2 mcs the conditional dynamics of field statistical properties, corresponded to ground state detection is shown (solid line). The probability of this event is slowing down due to the presence of real photons in the cavity and increasing the probability of atomic excitation. At the same time the conditional value of the photon creation rate decreases which is the consequence of the expected detection result. The quadrature dispersion in this time interval evaluates in the same way. It should be noted that corresponding dynamics in the absence of DCE field generation is rather different from the described above. Namely, due to the "switching off" of the process of photon creation (dashed line), the corresponding detection probability decreases more slowly and mean photon number becomes decreasing. The dispersion of quadrature tends to the "nonsqueezed" values.

After the first atom is detected in its ground state, the second atom goes into the cavity and begins interaction with the mode. This takes place at the third time interval from 2 to 3 mcs. All statistical properties behave in this case like in the previous one. It is worthwhile to recall, that presented dependences corresponding to exact and approximate solutions are quite different. And this means that for using the approximate results one should choose the interaction time small enough to save this approximation eligible.

In Fig.2 we show the probabilities to detect one (Fig.2a,b) and two (Fig.2c) subsequent atoms in their ground states. In Fig.2b we use additional transformation of the atomic state due to the interaction with the classical mode of EM field. Solid lines correspond to exact expressions for field transformers while dashed lines correspond to their expansion up to the second order. Dimensionless interaction time $\tau$ is chosen to be small compared to the time scale in the previous picture. So, all dependences are close to the origin.

From the presented probabilities, which may be defined from the statistics of the atomic state detection results, one can calculate the quadrature dispersion and other statistical characteristics of the monitored intracavity field. For instance we calculate here the mean photon number and the dispersion of $X$ quadrature after DCE field generation lasting 1 mcs. From here one can see that the state in fact is squeezed:

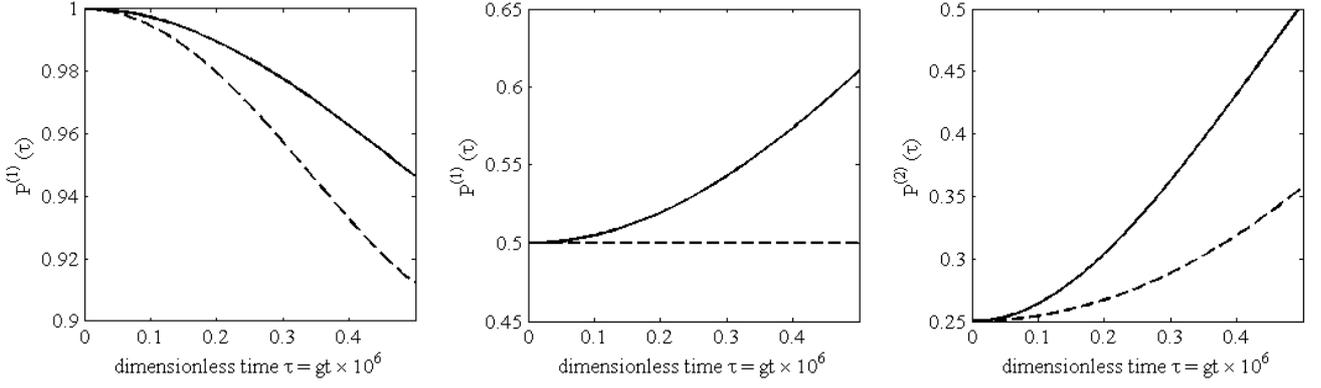

**Fig. 2** Probabilities to detect one (a, b) and two (c) subsequent atoms in their ground states; in (b) additional transformation of theatomic state is used. Solid lines correspond to exact expressions for field transformers while dashed lines correspond to series up to the second order.

$$\langle n \rangle = \frac{d^2}{d\tau^2} P_g^{(1)} = 0.28, \quad \langle X \rangle = \sqrt{2}\frac{d}{d\tau}\tilde{P}_g^{(1)}(0) = 0, \quad \langle X^2 \rangle = \frac{1}{2} - \frac{d^2}{d\tau^2}\tilde{P}_g^{(2)}(0) = 0.18. \qquad (27)$$

## 6. Concluding remarks

We proposed the general indirect measurement protocol for the intracavity mode generated in Dynamical Casimir Effect. Particularly, we concentrated on the statistical characteristics which confirm the nonthermal character of the monitored field. This way we present the algorithm for the squeezing detection without a homodyne scheme, which required additional amplification of the signal. Also we show how mean photon number may be extracted from atomic statistics of detectors clicks.

In our scheme we proposed that all atoms passing through the cavity are well separated in time, entrance into the cavity at certain moments of time and well selected by velocity. Also, just before each act of photodetection the cavity should be cleaned and prepared for the next measurement procedure. This may be done by sending the atoms one by one into the cavity and measuring them in the excited state. Thereby the approximate number of atoms required for realizing the considered measurement procedure is several tens of thousands with velocities of about thousands m/s.

For the following it would be interesting to estimate the influence of such factors as imperfectness of the atomic state detector and photon scattering on statistical properties due to the nonideality of the cavity walls. Also the more realistic case with probabilistic distribution of atoms velocities should be considered.

### Acknowledgements

This work was partially financially supported by the Government ofthe Russian Federation (grant 074-U01), by Ministry of Education andScience of the Russian Federation (GOSZADANIE 2014/190, Project14.Z50.31.0031 and ZADANIE No. 1.754.2014/К), by grant of Russian Foundation for BasicResearches and grant of the President of Russia (MK-2736.2015.2).